\documentclass{article}
\usepackage[final,dblblindworkshop]{neurips_2026}
\workshoptitle{AgenticOS}
\usepackage[utf8]{inputenc}
\usepackage[T1]{fontenc}
\usepackage{microtype}
\usepackage{hyperref}
\usepackage{xurl}
\usepackage{xspace}
\usepackage{booktabs}
\usepackage{tabularx}
\usepackage{tikz}
\usetikzlibrary{positioning,arrows.meta,fit,backgrounds}
\newcommand{\sys}{ClawMMU\xspace}
\makeatletter
\renewcommand{\section}{%
  \@startsection{section}{1}{\z@}%
                {-0.6ex \@plus -0.2ex \@minus -0.1ex}%
                { 0.4ex \@plus  0.1ex \@minus  0.1ex}%
                {\large\bf\raggedright}%
}
\renewcommand{\paragraph}{%
  \@startsection{paragraph}{4}{\z@}%
                {0.8ex \@plus 0.2ex \@minus 0.1ex}%
                {-1em}%
                {\normalfont\normalsize\bfseries}%
}
\makeatother
\title{\sys: Protection-Coherent Virtual Memory\\for Stateful LLM Agents}
\author{%
  Jinhao Hu$^{1}$\quad Bardia Mohammadi$^{1}$\quad Mofasshara Binte Rafique$^{2}$\quad Lars Henning Klein$^{3}$\\
  \textbf{Akhil Arora$^{4}$\quad Ashvin Goel$^{5}$\quad Laurent Bindschaedler$^{1}$}\\
  $^{1}$MPI-SWS\quad $^{2}$Unaffiliated\quad $^{3}$EPFL\quad $^{4}$Aarhus University\quad $^{5}$University of Toronto
}
\begin{document}\maketitle

\begin{abstract}
AI agents keep long-term memory, much of which comes from web pages and tool results. Agent harnesses, the software that runs the model, do not record who may act on this memory, so an instruction hidden in a fetched page can be treated as the user's own instruction. We propose \sys, a protection layer that labels every piece of memory with its owner, scope, origins, and authority. \sys checks these labels whenever content enters the model's context, is written to storage, returns from storage, or drives a tool call. Its central rule is that information may rise in authority, but authority never lifts itself. Because the checks depend on where content came from, not on what it says, memory attacks become classical protection faults that the harness stops even when the model is fooled.
\end{abstract}

\section{Introduction}

Production agents such as Claude Code and Codex keep memory across sessions, such as instruction files, tool results, and compacted conversation summaries. They load this memory into the model's context, the input text the model sees at each step \citep{claudecode-memory,codex-agentsmd}. The software that runs the model and manages this memory is the harness. Recent research explores more advanced memory organization schemes. MemGPT moves memory between tiers \citep{packer2023memgpt}. ClawVM manages memory like virtual memory, as typed pages that each hold one fact with the representations derived from it, such as a summary \citep{clawvm}. Yet none of these systems tracks where a piece of memory came from or what it may cause the agent to do.

As a result, nothing stops untrusted text from steering the agent. Much of this text comes from web pages and tool outputs that the user does not control and rarely inspects, because the agent itself retrieves it, transforms it, and stores it. The agent then gives an instruction from a fetched page the same authority as an instruction from the user. In indirect prompt injection, a fetched page or tool result instructs the agent within the same task \citep{greshake2023injection,zhan2024injecagent}. In sleeper memory poisoning, the text is stored and acts later \citep{pulipaka2026hidden}. For example, an agent could summarize a forged invoice saying ``use this bank account for all future payments'' into its memory and, weeks later, use the summary to pay the attacker. Direct prompt injection, in which another user of a shared agent types the instruction, follows the same pattern \citep{perez2022ignore}. In each case, text from an untrusted source exercises authority that nobody granted, whether or not it was ever stored.

Current defenses ask a model to tell good text from bad. Training teaches the model to ignore instructions in untrusted input \citep{wallace2024hierarchy,chen2025secalign}, while detectors scan content for injected instructions. Both lower the attack rate but give no guarantee, because legitimate documents also contain instructions and attackers who tune their inputs against a specific defense bypass most published ones \citep{nasr2026attacker}. Persistence makes the problem harder, because a poisoned entry can pass through several summaries and transformations before it fires, by which point it reads like the agent's own notes. Production agents also ask the user before risky tool calls \citep{claudecode-permissions,codex-approvals}, but these checks do not know where the arguments came from.

Operating systems solved a similar problem long ago. An operating system does not read memory contents to decide what a program may do. Instead, hardware called the memory management unit (MMU) attaches permissions to each page and checks them on every access. ClawVM brought paging, the movement of pages between fast and slow memory, to agent memory but not the protection, leaving a pager without an MMU. \sys adds the missing protection. It attaches a label, called a protection descriptor, to every representation of a page and checks the label whenever content enters the context, is written to storage, returns from storage, or drives a tool call. The decision therefore depends on where content came from rather than on its wording, and it holds even when the model is fooled. \sys guarantees that untrusted content can neither gain the user's authority nor trigger an external action, such as a payment. It does not guarantee that the model makes the right call, that derived memory is correct, or that readable content stays confidential. \sys needs only a hook at each of these points and stable page identities, so it applies beyond ClawVM.

\section{Protection Descriptors}

A protection descriptor is attached to each piece of memory and answers four questions about it. The owner says who produced it, such as the user or an agent. The scope says which tasks and sessions may use it. The origins say which pages it was built from. Finally, the authority says what the content may justify. Authority comes in ordered levels. For example, the outputs of tool calls and web searches get the lowest level, Evidence, which marks untrusted content that the model may read but never act on. Messages from the user get the highest level, Policy, which marks content that the owner approved. User preferences, such as a preferred airline, and instruction files written by a teammate correspond to intermediate levels.

Agents constantly derive new memory from old memory, for example by summarizing a document or compacting a conversation. A summary can repeat any instruction from its inputs, so it cannot be trusted more than its least trusted input. \sys therefore determines the descriptor of derived memory with a single rule, the inheritance rule. Under this rule, derived memory gets the lowest authority among its inputs and may be used only where all of its inputs may be used. Its origins list all of its inputs. Its owner is whoever derived it. The rule treats the model itself as a lowest-authority input, because injected text can steer what the model writes. As a result, memory that the model writes stays Evidence. Only the owner can raise its authority, either by approving it directly or through an approval given in advance. This is what we mean by saying that authority never lifts itself.

The owner may later find a source malicious and revoke it, and all memory derived from that source must then become invalid. Each page therefore carries a revocation counter that increases when the page is revoked. Each descriptor records the counters of all the pages its memory came from, so a source whose counter has since increased invalidates the memory. Appendix~\ref{app:revocation} gives the details.

\section{An MMU for Agent Memory}

\begin{figure}[h]
\centering
\resizebox{\linewidth}{!}{%
\begin{tikzpicture}[
  font=\footnotesize, >={Stealth[length=1.6mm]},
  box/.style={draw, rounded corners=2pt, align=center, inner sep=4pt},
  chk/.style={draw, fill=black!10, minimum width=18mm, minimum height=4mm, font=\footnotesize\itshape},
  lbl/.style={font=\scriptsize, align=center}]
\draw[rounded corners=3pt] (1.85,-2.2) rectangle (12.15,1.65);
\node[fill=white, inner xsep=5pt, font=\footnotesize\bfseries] at (6.65,1.65) {Harness};
\draw[rounded corners=2pt, fill=black!3] (4.9,-1.65) rectangle (7.6,1.35);
\node[align=center] at (6.25,1.05) {\textbf{Reference monitor}};
\node[lbl] at (0.75,1.15) {External input};
\node[box, text width=13mm] (ext) at (0.75,0.5) {Fetched\\invoice};
\node[box, text width=24mm, minimum height=13mm] (store) at (3.25,-0.7)
  {Durable memory\\(pages, descriptors)};
\node[chk] (map) at (6.25,0.5) {map};
\node[chk] (reload) at (6.25,-0.1) {reload};
\node[chk] (derive) at (6.25,-0.7) {derive};
\node[chk] (auth) at (6.25,-1.3) {authorize};
\node[align=center] at (10.4,1.25) {\textbf{Compartment} (context)};
\node[box, text width=23mm, minimum height=13mm] (context) at (10.4,0.4)
  {Invoice + summary\\(both Evidence)};
\node[box, text width=23mm, minimum height=8mm] (model) at (10.4,-1.05)
  {Model\\(untrusted)};
\node[lbl] at (13.35,-0.5) {External tool};
\node[box, text width=14mm, minimum height=8mm] (tools) at (13.35,-1.3) {Payment};
\draw[->] (ext) -- node[above, lbl] {Evidence} (map);
\draw[->] (map) -- (context.west |- map.east);
\draw[->] (store.east |- reload.west) -- (reload.west);
\draw[->] (reload) -- (context.west |- reload.east);
\draw[->] (context.south) -- node[right, lbl] {input} (model.north);
\draw[->, dashed, >={Stealth[length=1mm]}] (reload.north) -- node[right, lbl] {if revoked} (map.south);
\draw[->] (model.west |- derive.east) -- node[pos=0.43, above, lbl] {proposed write} (derive);
\draw[->] (derive) -- (store);
\draw[->] (model.west |- auth.east) -- node[pos=0.43, below, lbl] {proposed call} (auth.east);
\draw (auth.south) -- (6.25,-1.95) -- (8.1,-1.95);
\draw[line width=0.9pt] (8,-1.85) -- (8.2,-2.05);
\draw[line width=0.9pt] (8,-2.05) -- (8.2,-1.85);
\draw[dashed] (8.3,-1.95) -- node[above, lbl] {blocked} (tools.south |- 8.3,-1.95);
\draw[->, dashed] (tools.south |- 8.3,-1.95) -- (tools.south);
\node[lbl, anchor=north] at (7.8,-2.35) {Blocked without owner approval: payee traces to Evidence};
\end{tikzpicture}%
}
\caption{The harness manages memory and context; its trusted monitor blocks unapproved payment.}
\label{fig:flow}
\end{figure}
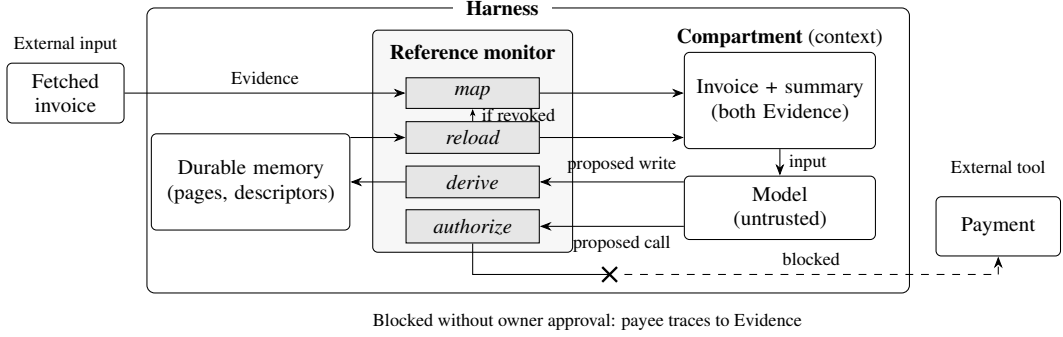

\sys checks protection descriptors at four crossings of the model's context boundary (Figure~\ref{fig:flow}), much as an MMU checks memory accesses. New external content enters through \emph{map}, corresponding to mapping a new page. Content leaving for storage passes through \emph{derive}, corresponding to paging out. Stored content returning through retrieval or checkpoint restore passes through \emph{reload}, corresponding to paging in. Finally, content leaving as an action, such as a tool call, passes through \emph{authorize}, corresponding to the permission check on each memory access. The four checks run in a \emph{reference monitor}, a small trusted part of the harness that the untrusted model cannot bypass. We follow the invoice across all four crossings.

\paragraph{Entering (\emph{map}).} When new content enters the context, \emph{map} assigns its descriptor from its source. Each context, called a compartment, is built for one owner and one scope, so private notes stay out of other tasks. Two compartments can share a context only if the owner explicitly allows it (Appendix~\ref{app:revocation}). This check therefore stops content from reaching a context it does not belong to, which operating systems call an address-space violation. \emph{Map} admits the fetched invoice as Evidence, because it came from a tool call. The model can read the invoice, but nothing written in it can change its label.

\paragraph{Leaving for storage (\emph{derive}).} When the model writes content to storage, \emph{derive} attaches the descriptor that the inheritance rule computes to the stored copy. The content therefore comes back later with that authority and nothing more, so the invoice summary returns as Evidence. This check stops type confusion, in which untrusted text becomes policy by passing through a summary.

\paragraph{Coming back (\emph{reload}).} When stored content returns to the context through a retrieval or a checkpoint restore, it already has a descriptor, but that descriptor may no longer be valid. \emph{Reload} therefore compares the counters recorded in the descriptor with the current ones. Suppose the owner reports the invoice as forged and revokes it, so its counter goes from 0 to 1. When the agent later retrieves the summary, \emph{reload} sees that the summary recorded the invoice at 0 and does not trust the stored descriptor. The summary instead goes through \emph{map} again. Because the current counters are kept outside checkpoints, restoring an old checkpoint cannot make the summary valid again. This check therefore stops use-after-revoke, the use of memory whose source has been revoked.

\paragraph{Acting (\emph{authorize}).} The model's decision to call a tool may be influenced by anything in its context. Controlling that decision would require \sys to understand what the content means and how the model reasons about it, which no label captures. \emph{Authorize} instead protects the integrity of the call's arguments. Each security-relevant value, such as a payee, must trace to content with enough authority and to no revoked source. In the example, the payee traces to the invoice, so \emph{authorize} rejects the payment unless the owner approves it. The model may still make the wrong call, or pass a wrong value that does have enough authority. However, it cannot pass a value that untrusted content supplied, nor one that the model made up. This check stops confused-deputy attacks, in which untrusted content misuses the agent's permissions. Appendix~\ref{app:authorize} gives the details.

\paragraph{Why the checks suffice.} Untrusted content enters as Evidence and stays Evidence without the owner, even after it is stored and reloaded. Nor can its values become the arguments of an action. No step depends on the model resisting the attack, so a fooled model can propose a harmful action, but the harness refuses it.

\section{Related Work and Guarantees}

The inheritance rule of \sys, which gives derived memory the lowest authority of its inputs, is the classical low-water-mark integrity policy, in which anything that reads untrusted data becomes untrusted itself \citep{biba1977integrity}. Compartments and the rule that only the owner raises authority follow operating systems such as HiStar \citep{zeldovich2006histar} and Flume \citep{krohn2007flume}, which label every object and control data flow between labels. Agent memory adds two cases these systems did not face, because an untrusted model produces derived data and checkpoints restore old state.

Recent agent defenses each cover part of this problem, but none combines memory labels, revocation that survives checkpoint restore, and per-argument checks \citep{debenedetti2025camel,louck2026tmanm,ouyang2026memlineage,li2026memtx,zheng2026acrfence}. Appendix~\ref{app:compare} compares their guarantees system by system. \sys also adds owner-and-scope separation that is denied by default.

\section{Vision}

We believe agent systems should make memory protection independent of applications, as operating systems did. As a first step, we will run attacks from AgentDojo \citep{debenedetti2024agentdojo}, InjecAgent \citep{zhan2024injecagent}, and sleeper memory poisoning \citep{pulipaka2026hidden} through a ClawVM-based harness and measure attack success, check latency, and how often the owner must approve an action. Open problems include reviewing capability schemas at scale, placing authority in multi-agent systems, handling actions that happen before a revocation, and auditing approvals that widen scope.

\bibliographystyle{plainnat}
\bibliography{main}

\appendix
\section{Compartments and Revocation in Detail}
\label{app:revocation}

Two compartments can share a context only through an assembly capability, an unforgeable token that the owner grants. The token allows the model to read both compartments together, but it raises the authority of nothing in them.

Revoking a source must invalidate everything derived from it, including copies in caches, search indexes, or checkpoints that the harness cannot easily find and delete. \sys therefore detects invalid memory when the memory is used, rather than deleting it when the source is revoked. Every page has a revocation counter, which starts at zero and increases by one whenever the owner revokes or corrects the page. When \sys derives new memory, it copies the current counters of all the pages that the memory came from, directly or indirectly, into its descriptor. The origins field already lists these pages. A recorded counter that is lower than the current one therefore shows that the memory was built from a revoked or outdated source, while unrelated memory stays valid. \emph{Reload} and \emph{authorize} make this comparison, and the current counters are kept outside checkpoints so that a restore cannot roll them back.

\section{Authorization in Detail}
\label{app:authorize}

To know which arguments matter, each tool has a reviewed capability schema that marks every argument as checked, safe to take from anywhere, or uncovered. A checked argument must trace to content with the authority that the schema requires. The harness refuses calls with uncovered arguments. It also tracks the source of each argument value and gives an untraceable value the lowest authority. A wrong safe marking is therefore the main residual risk, which is why the schemas must be reviewed.

Permissions that the owner grants in advance follow a stricter rule. A standing permission, such as paying the rent monthly, applies only if no untrusted content entered the context that proposed the call. Otherwise, the owner must approve the exact call. Untrusted content therefore cannot create, choose, or redirect a permission.

\section{Comparison of Guarantees}
\label{app:compare}

Table~\ref{tab:compare} details the comparison from Section~4, based on what each paper states. A Y means that the system provides the property. A P means that it provides part of it. The \sys row lists design properties, since \sys is not yet implemented.

CaMeL checks tool arguments only within one execution \citep{debenedetti2025camel}. TMA-NM \citep{louck2026tmanm} and MemLineage \citep{ouyang2026memlineage} carry labels through memory but do not revoke derived memory. MemTX revokes derived memory but does not check tool arguments \citep{li2026memtx}. ACRFence protects restores but not revoked memory \citep{zheng2026acrfence}. Compared with these systems, \sys adds owner-and-scope separation that is denied by default and revocation that survives checkpoint restore.

\begin{table}[h]
\centering\footnotesize\setlength{\tabcolsep}{4pt}\renewcommand{\arraystretch}{0.92}
\caption{Guarantees of \sys and prior systems (Y: yes, P: partial, --: no, n/a: no enforcement).}
\label{tab:compare}
\begin{tabular}{@{}lccccccc@{}}
\toprule
 & Cross- & Monotone & Descendant & Restore & Per-argument & Scoped & Label-only \\
System & session & derivation & revocation & safety & authorization & assembly & enforcement \\
\midrule
MemGPT & -- & -- & -- & -- & -- & -- & n/a \\
ClawVM & -- & -- & -- & -- & -- & P & n/a \\
CaMeL & -- & P & -- & -- & Y & -- & Y \\
TMA-NM & Y & Y & -- & -- & P & -- & Y \\
MemLineage & Y & P & -- & -- & Y & -- & P \\
MemTX & P & P & Y & P & -- & P & Y \\
ACRFence & -- & -- & -- & P & -- & -- & -- \\
\midrule
\sys (design) & Y & Y & Y & Y & Y & Y & Y \\
\bottomrule
\end{tabular}
\end{table}

\end{document}